\documentclass[journal]{IEEEtran}
\usepackage[utf8]{inputenc}
\usepackage{amsmath,amsfonts}
\usepackage{algorithmic}
\usepackage{algorithm}
\usepackage{array}
\usepackage[caption=false,font=normalsize,labelfont=sf,textfont=sf]{subfig}
\usepackage{textcomp}
\usepackage{stfloats}
\usepackage{url}
\usepackage{verbatim}
\usepackage{adjustbox}
\usepackage{graphicx}
\usepackage{cite}
\usepackage{balance}
\usepackage{booktabs}
\usepackage{makecell}
\usepackage{multirow}
\usepackage[normalem]{ulem} 
\usepackage{xcolor}
\usepackage{soul}
\soulregister\cite7 
\soulregister\ref7 
\usepackage{amsmath} 
\usepackage{booktabs}
\usepackage{makecell}
\usepackage{adjustbox}
\usepackage[markup=underlined,authormarkup=none,commandnameprefix=ifneeded]{changes}
\usepackage{color}
\usepackage{soul}

\begin{document}
\bstctlcite{IEEEexample:BSTcontrol}

\title{Bimorph Lithium Niobate Thickness-Shear Overtone Film Bulk Acoustic Resonator }

\author{Ziqian Yao,~\IEEEmembership{Student Member,~IEEE},
Ian Anderson,~\IEEEmembership{Student Member,~IEEE},
Tzu-Hsuan Hsu,~\IEEEmembership{Member,~IEEE},
Vakhtang Chulukhadze,~\IEEEmembership{Student Member,~IEEE},
Jack Kramer,~\IEEEmembership{Member,~IEEE},
\\and Ruochen Lu,~\IEEEmembership{Senior Member,~IEEE}%

\thanks{This work was supported by the National Science Foundation (NSF) under CAREER Award No. 2339731 and ECCS Award No. 2530883. Z. Yao was also supported by the NSF Research Traineeship (NRT) Quantum Cross-platform Advanced Training (QCAT) Fellowship. \textit{(Corresponding author: Ziqian Yao.)}}%
\thanks{Ziqian Yao, Ian Anderson, Tzu-Hsuan Hsu, Vakhtang Chulukhadze, and Ruochen Lu are with the Department of Electrical and Computer Engineering, The University of Texas at Austin, Austin, TX 78712 USA (e-mail: hanson.yao@utexas.edu).}%
\thanks{Jack Kramer was with the Department of Electrical and Computer Engineering, The University of Texas at Austin, Austin, TX 78712 USA. He is now with the Department of Electrical Engineering and Computer Sciences, University of California, Berkeley, Berkeley, CA 94720 USA.}%

}



\maketitle

\begin{abstract}

High quality factor ($Q$) is desirable in acoustic devices because it reduces resonance linewidth and benefits low-phase-noise oscillation, frequency control, and precision resonant sensing. Overtone operation further enables multiple discrete frequencies to be accessed from the same acoustic cavity without changing its physical dimensions, providing a compact route to frequency scalability. However, the main challenge is to preserve both high $Q$ and sufficient electromechanical coupling $k^2$ as the mode order increases. To address this tradeoff, we employ a bimorph periodically poled piezoelectric film (P3F) lithium niobate (LN) platform that combines the strong intrinsic piezoelectric response of X-cut LN with thickness-shear (TS) transduction for high-order overtone excitation. The device consists of a bonded $80~\mu\mathrm{m}$-thick single-crystal X-cut LN bimorph with opposite polarizations, patterned top and floating bottom electrodes, and a suspended air cavity. The P3F configuration mitigates charge cancellation associated with the alternating stress distribution of higher-order TS modes, enabling measurable electromechanical coupling across a broad sequence of overtones, while the thick LN acoustic cavity and increasingly confined high-order mode profiles support low-loss operation. The fabricated device exhibits TS overtones extending to 1.75~GHz. At room temperature, representative overtones at 0.77 and 0.89~GHz achieve 3-dB $Q$ values of 11,338 and 11,917, corresponding to $f\!\cdot\!Q$ products of $8.74\times10^{12}$ and $1.06\times10^{13}$~Hz, respectively. Temperature-dependent characterization from 297 to 12~K reveals systematic enhancement in $Q$, reaching a peak 3-dB $Q$ of 20,507 at 779~MHz and a maximum $f\!\cdot\!Q$ product of $1.98\times10^{13}$~Hz at 1.379~GHz. These results demonstrate bimorph P3F LN as a promising platform for high-$Q$, frequency-scalable micro-acoustic resonators in the sub-GHz and low-GHz regimes.

\end{abstract}


\section{Introduction}

High quality factor ($Q$) acoustic resonators are essential components for low-phase-noise oscillators, frequency control, and precision resonant sensing \cite{nelson201122muw,kourani2018tunable,tu2020dissipation}. In the sub-GHz and low-GHz regimes, high $Q$ provides a narrow resonance linewidth that benefits oscillator phase-noise performance and enables accurate detection of minute resonance-frequency shifts. Resonators supporting several distinct modes within a single device can further provide multiple operating frequencies while reducing resonator count and footprint, enabling new functionality such as self-temperature sensing or compensation \cite{zuo2011dual,kourani2020wideband,jabbari2025multi,zhang2025dual,vig2001temperature}.

Among the approaches for realizing this multi-frequency operation, overtone excitation is particularly attractive because multiple discrete resonances can be accessed through different mode orders without changing the device thickness. High-overtone bulk acoustic resonators (HBARs) have demonstrated high-$Q$ overtone operation across a range of piezoelectric and substrate platforms, from early microwave HBAR implementations \cite{haynes1985stable,lakin1993high,bailey1992frequency,zhang2006high} to more recent gallium nitride (GaN)- \cite{gokhale2020epitaxial}, lithium niobate (LN)- \cite{wu2021new,zhang2026impedance}, aluminum scandium nitride (AlScN)- \cite{gokhale2023x}, and Ga$_2$O$_3$-based platforms \cite{fu2025high}. Lateral-overtone bulk acoustic resonators (LOBARs) provide a complementary approach, with representative demonstrations in silicon carbide (SiC) \cite{gong2012ghz,jiang2021semi}, sapphire \cite{kuo2012micromachined}, and LN \cite{lu_lithium_2018,lu20205}. In oscillator systems, these multimode resonators can provide multiple operating frequencies from a single acoustic device, reducing resonator count and footprint \cite{yu2009hbar,kourani2020wideband,lu_lithium_2018,li2016resonance}.

The main challenge for practical overtone resonators is maintaining sufficient electromechanical coupling ($k^2$) while preserving high $Q$ at increasing mode order \cite{ziaei2011silicon,li2016resonance,lu_lithium_2018,kourani2020wideband}. In many overtone resonators, coupling to the acoustic cavity modes is relatively weak and generally decreases for higher-order tones as the transducer, designed for multimode operation, couples less effectively to the increasingly complex strain distribution of individual modes \cite{kourani2020wideband,lu_lithium_2018,li2016resonance}. The resulting reduction in $k^2$ lowers the combined figure of merit (FoM) = $Q\cdotp k^2$, increases motional impedance, and narrows the inductive region of the resonance, which can make oscillator implementation increasingly difficult despite high $Q$ \cite{li2016resonance,kourani2020wideband,zhang2026impedance}. A useful overtone resonator platform therefore requires a strongly piezoelectric material system capable of maintaining both high $Q$ and sufficient $k^2$ across multiple high-order modes \cite{wu2021new,gokhale2023x}.

To address these limitations, we demonstrate a high-$Q$ X-cut LN thickness-shear (TS) film bulk acoustic resonator (FBAR) based on a bimorph periodically poled piezoelectric film (P3F) structure. The resonator consists of two oppositely oriented X-cut LN layers, which mitigate charge cancellation, effectively double the acoustic film thickness without sacrificing $k^2$, and enable operation at twice the frequency of a conventional single-layer LN resonator with the same total film thickness. The resulting bimorph LN TS-FBAR supports a series of high-order TS overtones extending to 1.75~GHz, while the strong intrinsic piezoelectric response of X-cut LN provides a favorable basis for maintaining $k^2$ across multiple overtone modes in a thick and mechanically robust single-crystal acoustic cavity. At room temperature, representative overtones at 0.771 and 0.89~GHz exhibit 3-dB $Q$ values of 11,338 and 11,917, corresponding to $f\!\cdot\!Q$ products of $8.74\times10^{12}$ and $1.06\times10^{13}$~Hz, respectively. To investigate the loss mechanisms limiting $Q$, we perform temperature-dependent characterization from 297 to 12~K. A systematic enhancement in $Q$ is observed with decreasing temperature, reaching a peak 3-dB $Q$ of 20,507 at 779~MHz and a maximum $f\!\cdot\!Q$ product of $1.98\times10^{13}$~Hz at 1.379~GHz and 12~K, demonstrating that P3F-enabled high-order TS excitation in thick transferred LN provides a promising pathway toward compact, frequency-scalable, and low-loss micro-acoustic resonators.

\section{Device Design and Simulation}\label{sec2}


\begin{figure}[!t]
\centering
\includegraphics[width=3.5in]{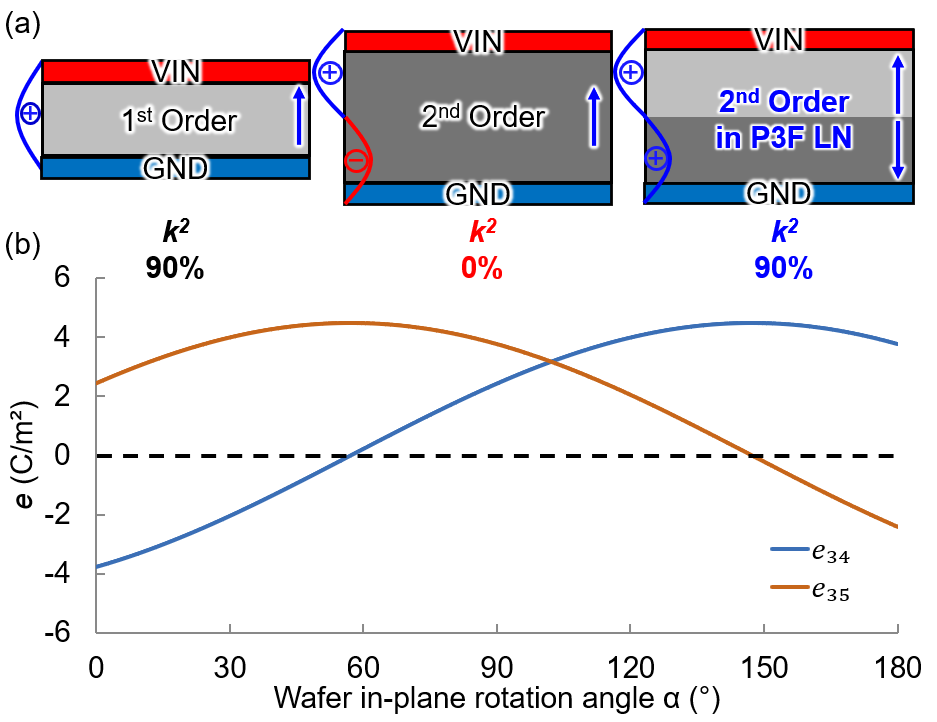}
\caption{Crystal orientation and P3F transduction principle of bimorph LN TS-FBAR. (a) Schematic of electrode and polarization configurations for conventional and P3F thickness-shear excitation. (b) Calculated $e_{34}$ and $e_{35}$ of X-cut LN as functions of wafer in-plane rotation angle $\alpha$.}
\label{fig_orientation}
\label{fig_2}
\end{figure}

X-cut LN is selected for its strong intrinsic piezoelectric coupling to TS motion. As seen in Fig.~\ref{fig_2}(a), a single X-cut LN layer can strongly excite the fundamental TS mode, with electrodes on opposite sides of the piezoelectric layer establishing an electric field primarily along the thickness direction. In X-cut LN, the thickness-directed electric field $E_3$ couples through $e_{34}$ and $e_{35}$ to generate the two shear-stress components $T_4$ and $T_5$. These shear components do not act independently, but are coupled by the anisotropic material response to form the fast- and slow-shear waves \cite{pijolat2009large}. Fig.~\ref{fig_2}(b) shows the calculated $e_{34}$ and $e_{35}$ as functions of the wafer in-plane rotation angle $\alpha$, illustrating the strong orientation dependence in X-cut LN.

To enable efficient excitation of even-order and higher-order TS modes, the two X-cut LN layers are arranged in a configuration with in-plane orientations of $\alpha=0^{\circ}$ and $180^{\circ}$, i.e., a P3F stack. In a uniformly oriented piezoelectric layer, the alternating stress distribution of the second-order TS mode leads to substantial charge cancellation and weak electromechanical transduction. A $180^{\circ}$ in-plane rotation reverses the signs of both $e_{34}$ and $e_{35}$ while preserving their magnitudes, producing opposite shear-transduction polarity between the two LN layers. As shown in Fig.~\ref{fig_2}(a), this polarity reversal compensates the stress-phase reversal of the second-order TS mode, allowing the piezoelectric contributions from the two layers to add constructively. Direct bonding of oppositely oriented X-cut lithium tantalate (LT) has previously been demonstrated for even-order TS excitation \cite{sugimoto1998even}, while more recent bimorph P3F X-cut LN structures have further demonstrated the effectiveness of orientation-engineered multilayer transduction \cite{chulukhadze2026bimorph,yao2025bimorph}, which enable the use of thicker LN films with a lower surface-to-volume ratio, reducing the impact of surface-induced loss. The same principle provides the basis for excitation of higher-order TS overtones, with the generalized multilayer P3F excitation mechanism described in \cite{kramer_acoustic_2025}.

\begin{figure}[!t]
\centering
\includegraphics[width=3.5in]{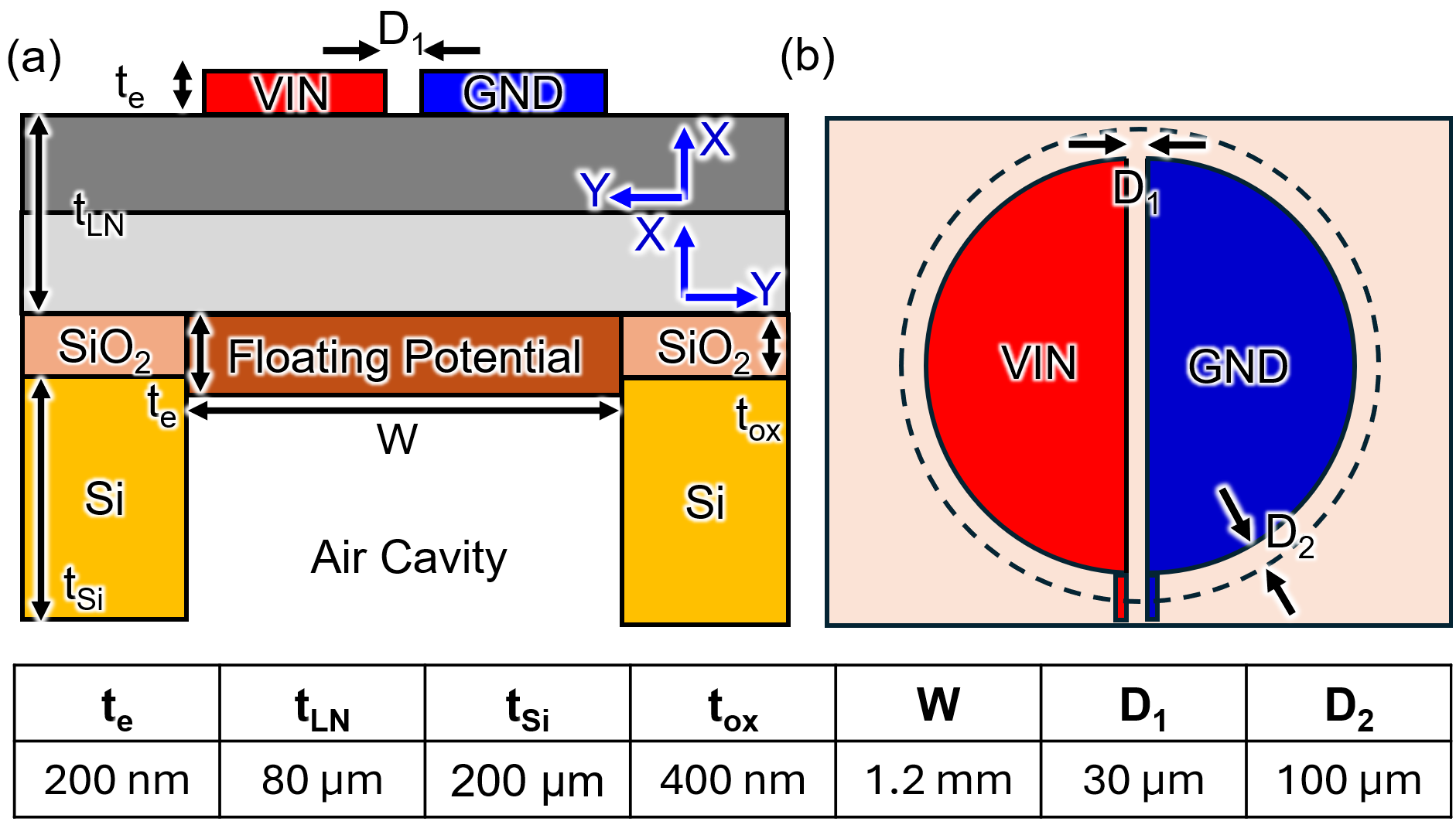}
\caption{Schematic and key dimensions of bimorph LN TS-FBAR. (a) Cross-sectional view showing the top electrodes, floating bottom electrode, bimorph LN stack, and suspended air cavity. (b) Top view showing VIN/GND electrodes and lateral dimensions. }
\label{fig_orientation}
\label{fig_3}
\end{figure}


The proposed P3F TS-FBAR is shown in Fig.~\ref{fig_3}. The device consists of an $80~\mu$m-thick bimorph X-cut P3F LN stack bonded to a silicon (Si) substrate with a silicon dioxide (SiO$_2$) interlayer. Patterned top electrodes are formed on the top surface of the LN stack, while a floating bottom electrode is located beneath the bottom LN and suspended over the backside air cavity. The top and bottom electrodes form a thickness-field excitation configuration and have closely matched lateral dimensions to maximize geometric overlap, promoting thickness-field acoustic energy confinement. Fig.~\ref{fig_3}(b) shows the top view of the resonator, where the dashed circle denotes the backside deep-Si-etch (DSE) region. A lateral clearance $D_2$ is introduced between the top-electrode edge and the DSE boundary to accommodate backside alignment tolerance during photolithography while maintaining sufficient overlap between electrodes, thereby reducing acoustic energy leakage near the cavity boundary. The SiO$_2$ layer serves as both an intermediate bonding layer and an etch-stop layer during the backside DSE process.

\begin{figure}[!t]
\centering
\includegraphics[width=3.5in]{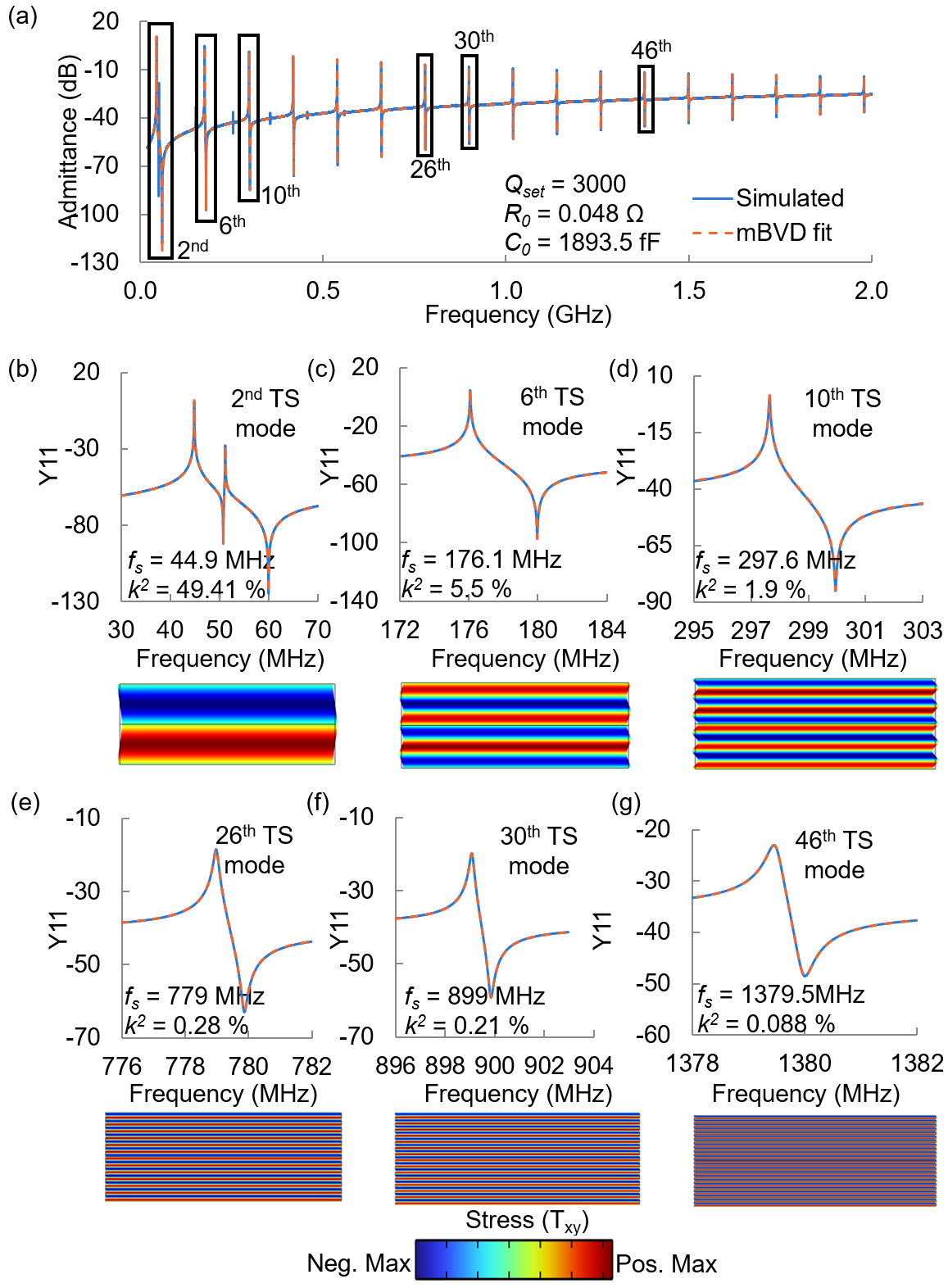}
\caption{Simulated admittance of bimorph P3F TS-FBAR. (a) Wideband admittance with mBVD fitting. (b)--(g) Enlarged admittance responses of 2nd-, 6th-, 10th-, 26th-, 30th-, and 46th-order TS modes, respectively, together with the corresponding cross-sectional shear-stress distributions.}
\label{fig_orientation}
\label{fig_4}
\end{figure}

To investigate the high-order TS overtone performance, a two-dimensional (2D) finite-element analysis (FEA) is performed in COMSOL. A mechanical $Q$ of 3000 is assigned in the simulation, and the resulting admittance response is fitted using a modified Butterworth--Van Dyke (mBVD) equivalent-circuit model [Fig.~\ref{fig_4}(a)]. The wideband response exhibits a series of TS overtones. As highlighted in Fig.~\ref{fig_4}(b)--(g), the 2nd-, 6th-, 10th-, 26th-, 30th-, and 46th-order TS modes exhibit series frequencies ($f_s$) of 44.9, 176.1, 297.6, 779, 899, and 1379.5~MHz, with corresponding $k^2$ values of 49.41\%, 5.5\%, 1.9\%, 0.28\%, 0.21\%, and 0.088\%, respectively. The additional resonance adjacent to the 2nd-order TS mode in Fig.~\ref{fig_4}(b) is attributed to the secondary slow-shear mode of X-cut LN \cite{bousquet2019single}, which is most prominent for this lowest-order tone because of its substantially larger $k^2$. The corresponding cross-sectional shear-stress ($T_{xy}$) distributions, evaluated in the device $x$--$z$ plane oriented $60^{\circ}$ with respect to the crystallographic $Y$ axis, show alternating stress lobes through the P3F LN thickness, with the number of lobes increasing with TS mode order. These results demonstrate that the P3F stack enables piezoelectric transduction of high-order even TS overtones while maintaining coupling over a broad frequency range.

\begin{figure}[!t]
\centering
\includegraphics[width=3.5in]{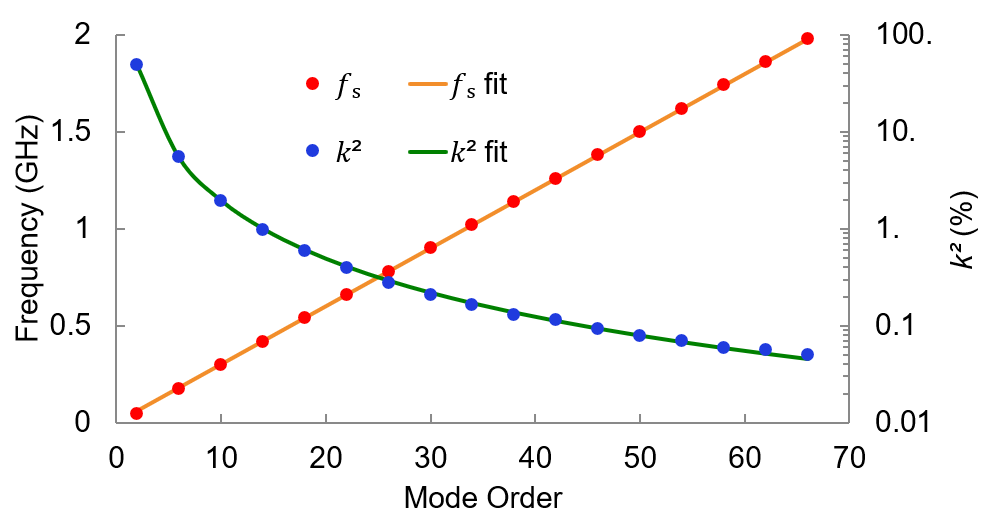}
\caption{Simulated $f_s$ and $k^2$ versus TS mode order with fitted scaling curves, where $N$ denotes the TS mode order. The resonance scales approximately as $f_s\propto N$, while $k^2$ follows an approximate $N^{-2}$ dependence.}
\label{fig_orientation}
\label{fig_6}
\end{figure}

Fig.~\ref{fig_6} plots the extracted $f_s$ and $k^2$ as functions of TS mode order in FEA. We fit the data to the expected thickness-mode scaling relations, $f_s=AN$ and $k^2=B/N^2$, where $N$ denotes the TS mode order and $A$ and $B$ are fitting coefficients for frequency and coupling, respectively. The $f_s$ data follow the linear fit closely, confirming the nearly proportional increase in resonance frequency with mode order. In contrast, the higher-order $k^2$ values approximately follow an inverse-square trend, reflecting the progressive reduction in electromechanical coupling as mode order increases.

\begin{figure}[!t]
\centering
\includegraphics[width=3.5in]{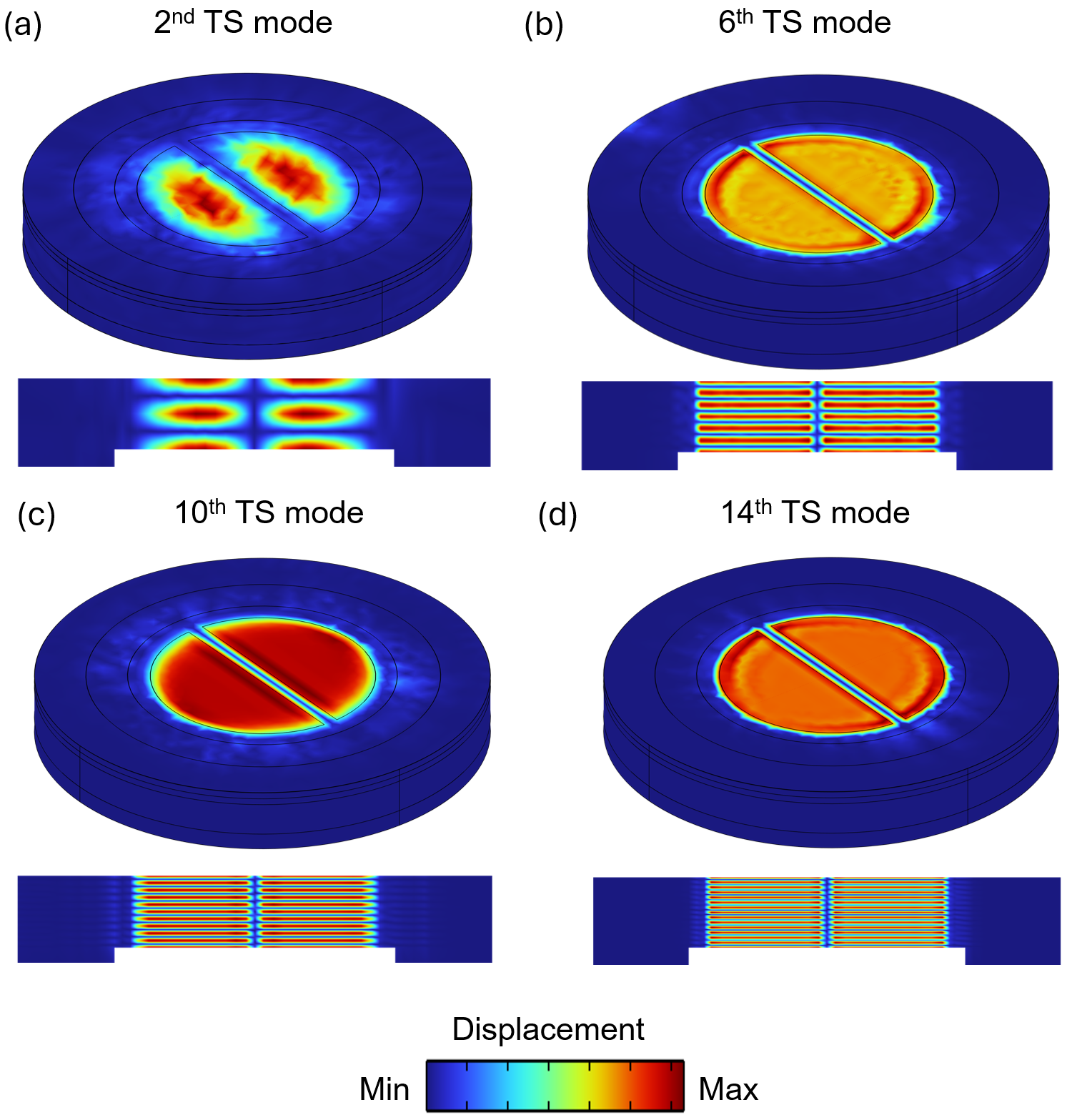}
\caption{3D FEA displacement distributions of bimorph P3F LN TS-FBAR at (a) 2nd-, (b) 6th-, (c) 10th-, and (d) 14th-order TS resonances. Each panel shows a three-dimensional view and the corresponding x–z cross-section.}
\label{fig_orientation}
\label{fig_5}
\end{figure}


To further examine the acoustic confinement of the TS overtones, three-dimensional (3D) FEA is performed for the 2nd-, 6th-, 10th-, and 14th-order TS modes. The model includes the bimorph LN stack, patterned electrodes, suspended membrane above an air cavity, and surrounding support terminated by perfectly matched layers (PMLs). No artificial mechanical damping is assigned in these 3D simulations, except for the outgoing elastic waves absorbed by PMLs. As shown in Fig.~\ref{fig_5}(a)--(d), the simulated displacement becomes progressively more localized within the electrode-covered suspended region as the TS mode order increases. The 2nd-order mode in Fig.~\ref{fig_5}(a) exhibits a relatively broad displacement distribution with noticeable extension toward the surrounding support, whereas the 6th-, 10th-, and 14th-order modes in Fig.~\ref{fig_5}(b)--(d) show increasingly well-defined lateral confinement within the suspended region. In particular, the 10th- and 14th-order modes exhibit highly localized displacement with only limited motion extending into the peripheral support. The corresponding cross-sectional profiles match those in 2D FEA. These results confirm effective lateral acoustic confinement for all four representative TS modes, with progressively stronger confinement at higher mode orders.

\section{Device Fabrication}\label{sec7}

\begin{figure}[!t]
\centering
\includegraphics[width=3.5in]{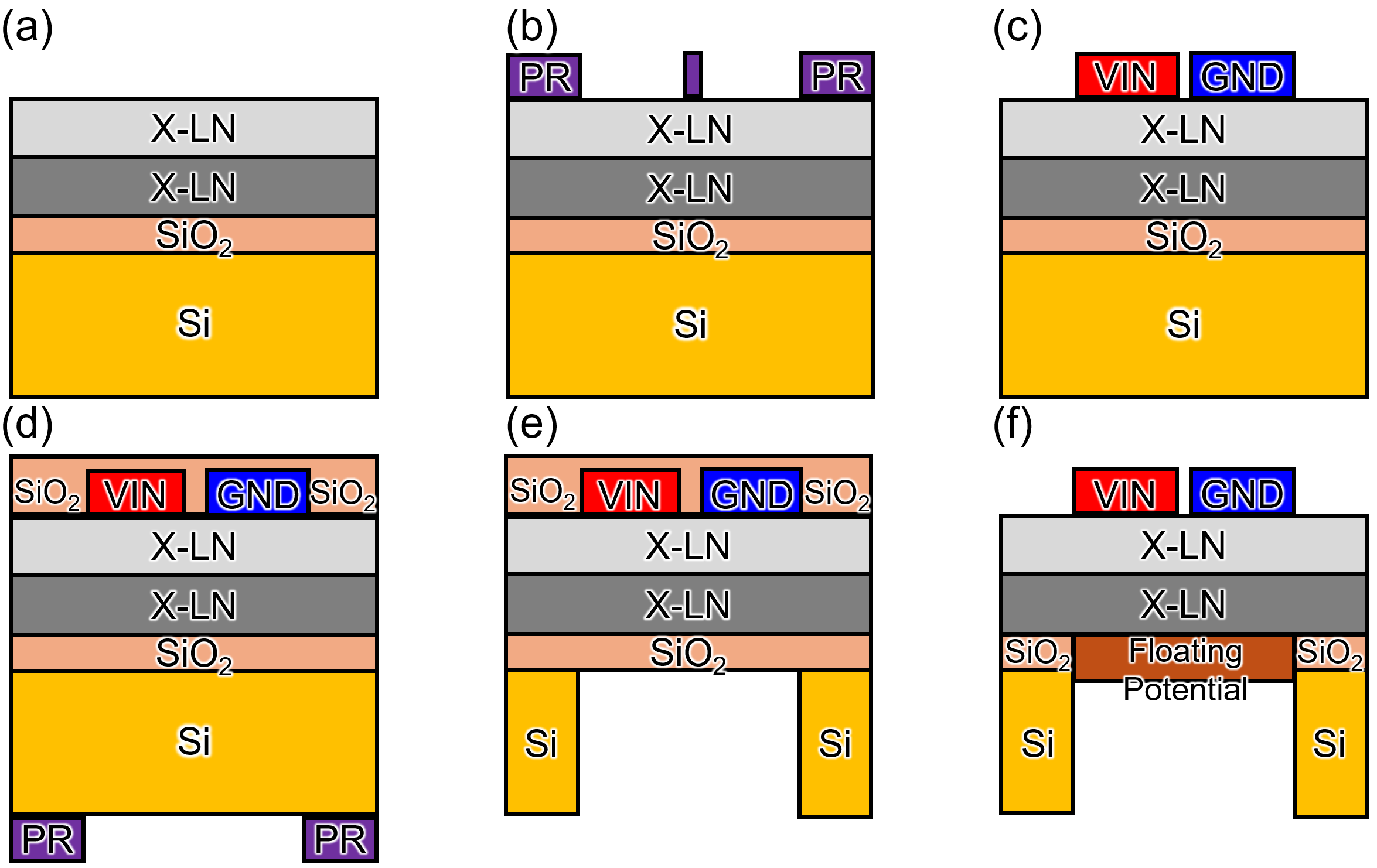}
\caption{Fabrication process flow. (a) Bonded bimorph X-cut LN/SiO$_2$/Si substrate. (b) Top-electrode patterning by photolithography. (c) Top electrode formation and lift-off. (d) PECVD SiO$_2$ protection of the top electrodes followed by backside-aligned lithography. (e) Backside deep-Si etching to release the LN membrane. (f) Removal of the protective and etch-stop SiO$_2$ layers followed by backside metal deposition to form the floating bottom electrode.}
\label{fig_orientation}
\label{fig_7}
\end{figure}

The fabrication process flow is shown in Fig.~\ref{fig_7}. The process starts from a bonded bimorph X-cut LN-on-Si wafer provided by NGK Corporation. The stack consists of two 40~$\mu$m-thick X-cut P3F LN layers and a 400-nm-thick SiO$_2$ interlayer on top of the 200~$\mu$m-thick Si (111) carrier wafer. Top-side photolithography first defines the signal and ground electrodes, followed by e-beam evaporation of 10-nm chromium (Cr)/200-nm gold (Au) and metal lift-off in acetone. Cr serves as the adhesion layer while being compatible with the subsequent buffered oxide etch (BOE) process. A protective SiO$_2$ layer is then deposited by plasma-enhanced chemical vapor deposition (PECVD) to prevent mechanical damage during backside processing. Backside-aligned photolithography then defines the cavity opening. DSE removes the Si carrier within the defined region until it reaches the SiO$_2$ etch-stop layer. After DSE, acetone strips the backside photoresist, and BOE removes the exposed SiO$_2$ etch-stop and top-side protective oxide to release the bimorph LN membrane. Finally, 10-nm Cr/200-nm Au is evaporated from the backside through the released cavity to form the floating bottom electrode. Matching the top and bottom electrode thicknesses minimizes asymmetric acoustic mass loading.

\begin{figure}[!t]
\centering
\includegraphics[width=3.5in]{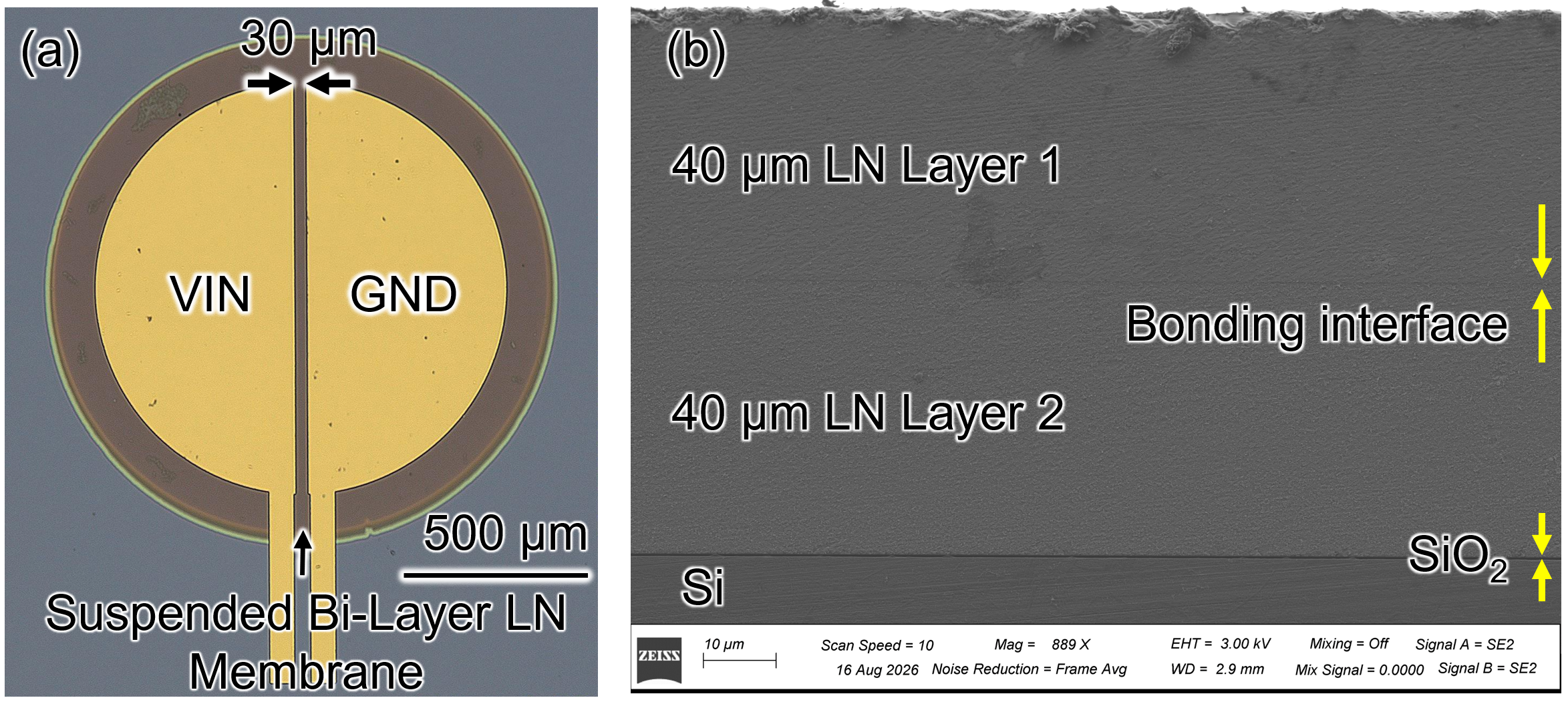}
\caption{(a) Optical image of the fabricated resonator showing the VIN/GND electrodes and suspended bilayer LN membrane. (b) Cross-sectional SEM image of the bonded device stack.}
\label{fig_orientation}
\label{fig_8}
\end{figure}

Fig.~\ref{fig_8} shows the optical and cross-sectional characterization of the fabricated device. The optical image in Fig.~\ref{fig_8}(a) shows clearly defined VIN and GND electrodes separated by a 30~$\mu$m gap and the released bilayer LN membrane suspended over the backside cavity. Fig.~\ref{fig_8}(b) presents a cross-sectional scanning electron microscope (SEM) image of the stack, where the two 40~$\mu$m-thick X-cut LN layers and the underlying Si substrate are clearly seen. These images confirm the successful formation of the bonded bimorph LN stack and suspended resonator structure.

\section{Measurement and Results}\label{sec8}

To characterize the wideband and temperature-dependent performance of the fabricated bimorph P3F TS-FBAR, we measure the device using a Lake Shore TTPX cryogenic probe station and a Keysight P5028A vector network analyzer (VNA). The sample is mounted to the cryogenic stage using copper tape to provide mechanical fixation and thermal contact, and the chamber is evacuated to approximately $5\times10^{-6}$~Torr. An initial measurement is performed at 297~K ambient laboratory temperature. Liquid helium is subsequently used to cool the system, and measurements are taken after the stage temperature stabilizes at each set point. 

The VNA measurements are performed with an input power of $-15$~dBm and an intermediate-frequency (IF) bandwidth of 1~kHz over a frequency range from 20~MHz to 1.75~GHz. A segmented-frequency sweep is employed, with finer frequency spacing in narrow windows around each TS resonance and coarser spacing between resonances. The frequency resolution around each resonance is selected to ensure at least 20 measurement points within the 3-dB bandwidth, providing sufficient sampling for reliable parameter extraction of $Q_s$.

\begin{figure}[!t]
\centering
\includegraphics[width=3.5in]{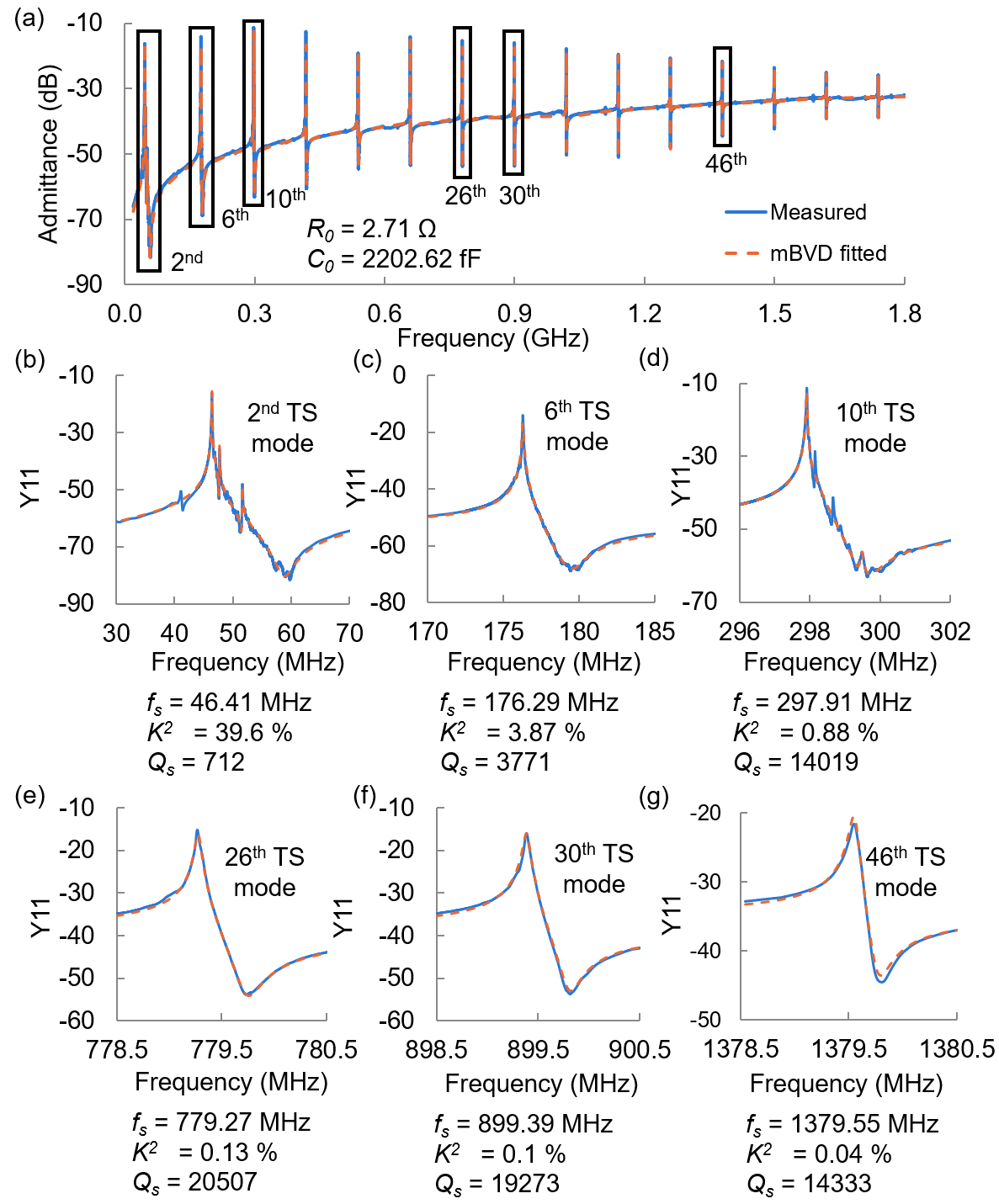}
\caption{Measured admittance of P3F TS-FBAR at 12~K. (a) Wideband response with mBVD fitting. (b)--(g) Enlarged admittance of 2nd-, 6th-, 10th-, 26th-, 30th-, and 46th-order TS modes, respectively.}
\label{fig_orientation}
\label{fig_9}
\end{figure}

Fig.~\ref{fig_9} summarizes the measured wideband response of the resonator at 12~K together with the corresponding mBVD fitting. As shown in Fig.~\ref{fig_9}(a), a series of TS overtones is observed. Fig.~\ref{fig_9}(b)--(d) show the 2nd-, 6th-, and 10th-order TS modes at $f_s=46.41$, 176.29, and 297.91~MHz, with corresponding $k^2$ values of 39.6\%, 3.87\%, and 0.88\% and 3-dB $Q_s$ values of 712, 3,771, and 14,019, respectively. At higher orders, the 26th- and 30th-order TS modes shown in Fig.~\ref{fig_9}(e) and (f) reach $Q_s$ values of 20,507 and 19,273 at 779.27 and 899.39~MHz, respectively, representing two of the highest measured $Q_s$ values. The 46th-order TS mode in Fig.~\ref{fig_9}(g) occurs at 1.38~GHz with $Q_s$ of 14,333 and yields the maximum measured $f\!\cdot\!Q$ product of approximately $1.98\times10^{13}$~Hz. The mBVD model closely fits the measured admittance, and the extracted mBVD parameters are summarized in Table~\ref{tab:measured_modes}.

\begin{table}[t]
\caption{Measured and mBVD-fitted parameters of TS overtones at 12~K}
\label{tab:measured_modes}
\centering

\footnotesize
\setlength{\tabcolsep}{2.0pt}
\renewcommand{\arraystretch}{1.08}

\begin{adjustbox}{width=0.99\columnwidth}
\begin{tabular}{@{}c c c c c c c c@{}}
\toprule
\makecell[c]{Mode\\Order} &
\makecell[c]{$f_s$\\(MHz)} &
\makecell[c]{$k^2$\\(\%)} &
\makecell[c]{3-dB\\$Q_s$} &
\makecell[c]{$f\!\cdot\!Q$\\(Hz)} &
\makecell[c]{$R_m$\\($\Omega$)} &
\makecell[c]{$L_m$\\($\mu$H)} &
\makecell[c]{$C_m$\\(fF)} \\
\midrule

2nd  & 46.41   & 39.60 & 712   & $3.30\times10^{10}$ & 6.14  & 16.64 & 706.97 \\
6th  & 176.29  & 3.87  & 3771  & $6.65\times10^{11}$ & 7.30  & 11.81 & 69.03  \\
10th & 297.91  & 0.88  & 14019 & $4.18\times10^{12}$ & 4.89  & 18.26 & 15.63  \\
14th & 418.51  & 0.44  & 12072 & $5.05\times10^{12}$ & 4.74  & 18.24 & 7.93   \\
18th & 538.84  & 0.23  & 6908  & $3.72\times10^{12}$ & 9.01  & 21.65 & 4.03   \\
22nd & 659.14  & 0.20  & 17195 & $1.13\times10^{13}$ & 5.98  & 16.12 & 3.62   \\
26th & 779.27  & 0.13  & 20507 & $1.60\times10^{13}$ & 6.37  & 18.25 & 2.29   \\
30th & 899.39  & 0.10  & 19273 & $1.73\times10^{13}$ & 6.53  & 17.50 & 1.79   \\
34th & 1019.46 & 0.08  & 16991 & $1.73\times10^{13}$ & 7.66  & 17.04 & 1.43   \\
38th & 1139.53 & 0.07  & 15194 & $1.73\times10^{13}$ & 9.20  & 16.11 & 1.21   \\
42nd & 1259.55 & 0.05  & 15267 & $1.92\times10^{13}$ & 10.68 & 17.68 & 0.90   \\
46th & 1379.55 & 0.04  & 14333 & $1.98\times10^{13}$ & 12.77 & 18.46 & 0.72   \\
50th & 1499.54 & 0.03  & 10906 & $1.64\times10^{13}$ & 15.55 & 19.95 & 0.56   \\
54th & 1619.49 & 0.03  & 8097  & $1.31\times10^{13}$ & 22.30 & 20.71 & 0.47   \\
58th & 1739.45 & 0.02  & 7248  & $1.26\times10^{13}$ & 22.30 & 20.99 & 0.40   \\

\bottomrule
\end{tabular}
\end{adjustbox}

\end{table}

\begin{figure}[!t]
\centering
\includegraphics[width=3.5in]{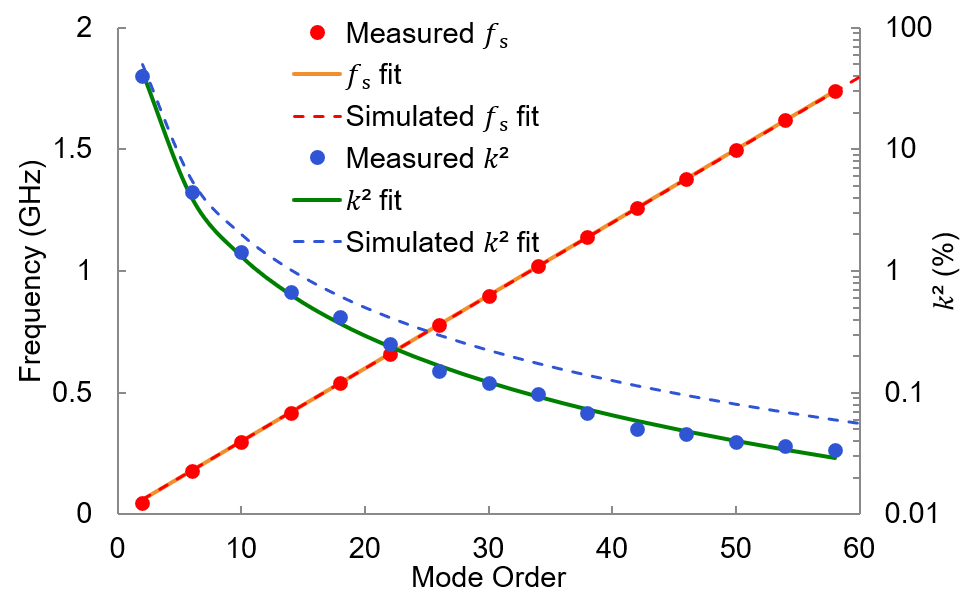}
\caption{Measured $f_s$ and $k^2$ of bimorph P3F TS-FBAR at 12~K as functions of TS mode order. Solid curves show fits to the measured data, while dashed curves show the corresponding simulated fits.}
\label{fig_orientation}
\label{fig_10}
\end{figure}

Fig.~\ref{fig_10} demonstrates that the measured TS overtone frequencies closely follow the simulated scaling, while a small systematic discrepancy remains in the $k^2$. The measured $f_s$ and $k^2$ are fitted as functions of TS mode order, with solid curves representing fits to the measured data and dashed curves representing the corresponding simulated fits. The measured $f_s$ closely follows the simulated linear trend across the entire overtone range. Similarly, both the measured and simulated $k^2$ exhibit a consistent reduction with increasing mode order and are well described by their respective fitted trends. However, the measured $k^2$ is slightly lower than the simulated prediction, which may arise from fabrication and device nonidealities not fully captured in the simulation. Nevertheless, both measurement and simulation exhibit the same overall reduction in $k^2$ with increasing mode order, confirming the coupling tradeoff associated with high-order TS excitation.

\begin{figure}[!t]
\centering
\includegraphics[width=3.5in]{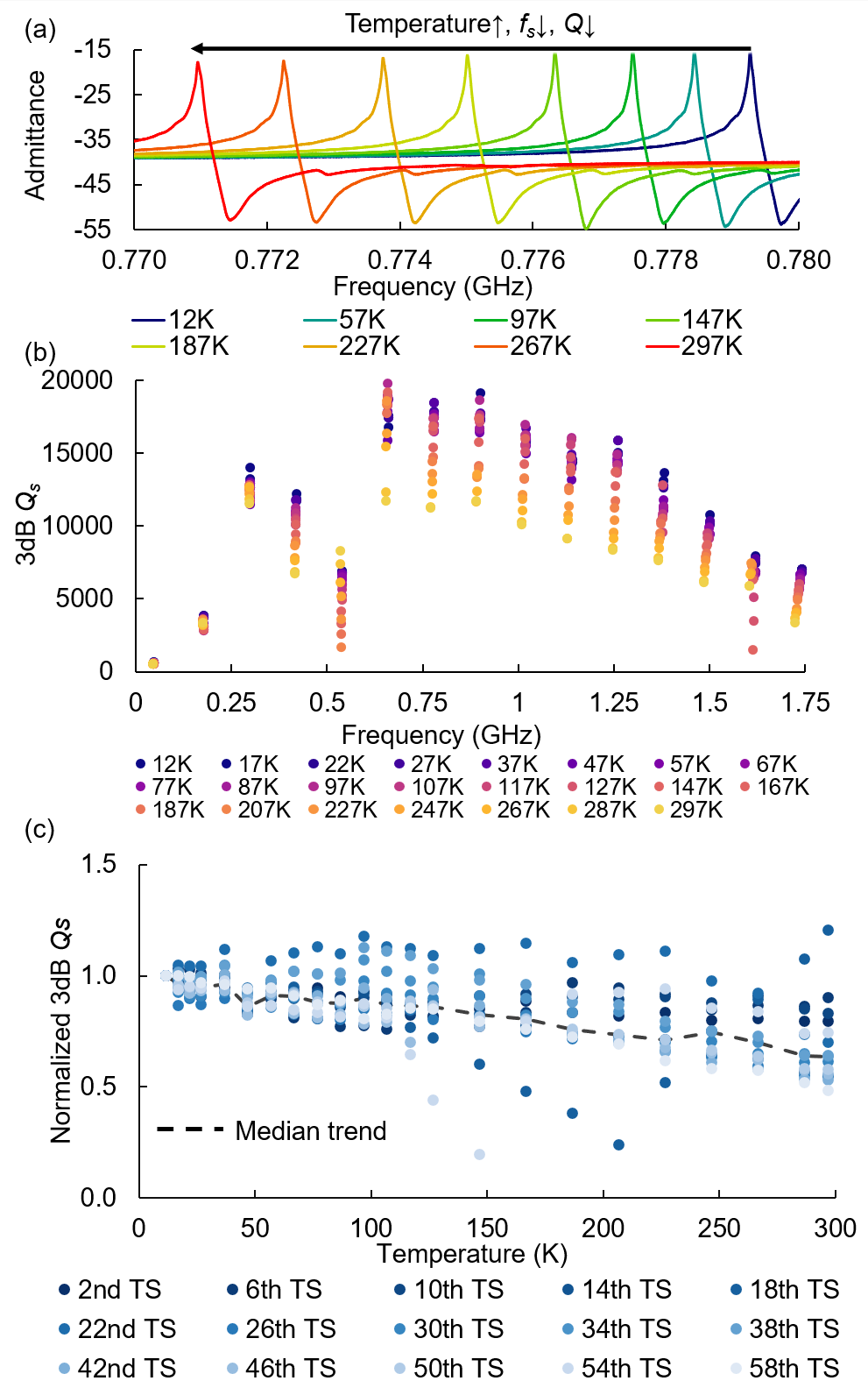}
\caption{Temperature-dependent characterization of bimorph P3F TS-FBAR from 297 to 12~K. (a) Measured admittance of the 26th-order TS mode, showing the temperature dependence of resonance frequency and 3-dB $Q_s$. (b) Extracted 3-dB $Q_s$ of multiple TS overtones over the full temperature range. (c) Corresponding 3-dB $Q_s$ normalized to their values at 12~K, with the dashed line indicating the median across all measured TS modes.}
\label{fig_orientation}
\label{fig_11}
\end{figure}

To investigate the loss mechanisms limiting the resonator $Q$, temperature-dependent measurements are performed from 297 to 12~K, as summarized in Fig.~\ref{fig_11}. Fig.~\ref{fig_11}(a) shows the measured admittance of the 26th-order TS mode, for which decreasing temperature produces both a shift in $f_s$ and a pronounced narrowing of the resonance bandwidth. Fig.~\ref{fig_11}(b) extends the analysis to the measured TS overtones and shows a general enhancement in $Q_s$ upon cooling, particularly for the principal high-$Q$ modes. Several lower-$Q$ outliers appear at frequencies where nearby spurious modes perturb the resonance response. To distinguish the relative temperature dependence among different modes, Fig.~\ref{fig_11}(c) plots the 3-dB $Q_s$ of each mode normalized to its value at 12~K, with the dashed line showing the median normalized $Q_s$ across all measured modes at each temperature. While most modes exhibit substantial $Q_s$ enhancement upon cooling, the normalized $Q_s$ of many modes tends to plateau at the lowest temperatures, indicating diminishing improvement with further cooling. This behavior suggests that, as the temperature-dependent intrinsic loss in LN is reduced, other loss contributions, such as surface/interface damping and anchor loss, may become increasingly important in limiting the measured $Q_s$. Quantitative separation of these loss mechanisms remains under investigation.

\begin{figure}[!t]
\centering
\includegraphics[width=3.5in]{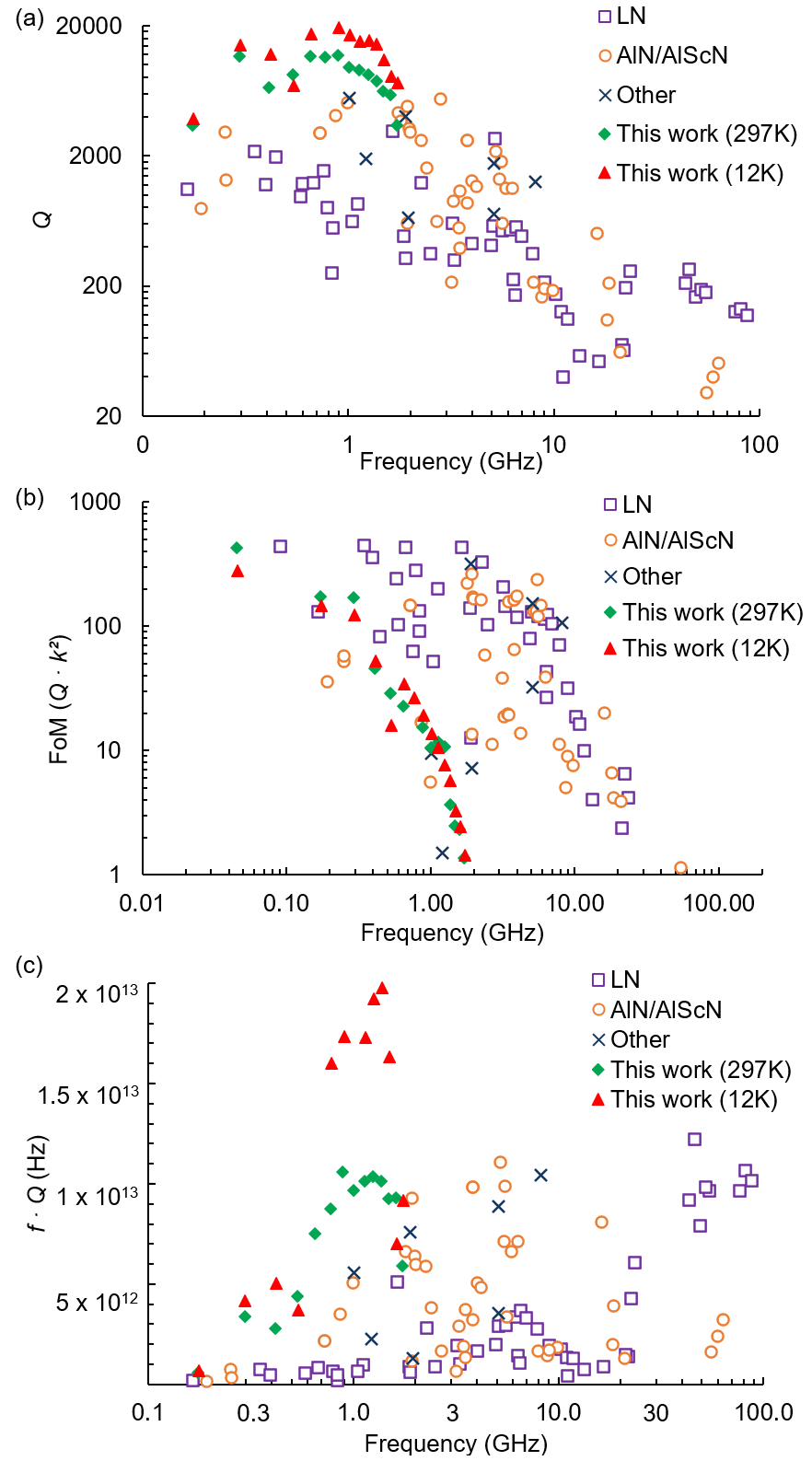}
\caption{State-of-the-art comparison of acoustic resonators  for which the targeted acoustic mode is primarily confined to the piezoelectric material or piezoelectric-guided acoustic stack. (a) Measured $Q$ versus operating frequency. (b) Corresponding figure of merit versus frequency. (c) $f\!\cdot\!Q$ product versus operating frequency.}
\label{fig_orientation}
\label{fig_12}
\end{figure}

Finally, we benchmark the measured performance against previously reported acoustic resonators in fully piezoelectric stacks in Fig.~\ref{fig_12}. The comparison spans representative resonator platforms, including surface acoustic wave (SAW), bulk acoustic wave (BAW), and laterally vibrating resonators (LVR). To provide a consistent basis for comparison, the benchmark focuses on resonators in which the piezoelectric material directly serves as the principal resonant acoustic medium. The LN dataset is compiled from \cite{colombo2018x,chen2019q,li2019temperature,olsson2014lamb,pop2018investigation,hsu2021thin,hsu2020large,yang2020high,zhang2020surface,lu2020a1,yang20194,zheng_near_2024,su_59_2023,kramer_acoustic_2025,campbell_52-ghz_2025,hsu2025toward,lee2025transverse,lee2025spectrum,campbell202421,gubinelli2025laterally}, while the AlN/AlScN dataset is compiled from \cite{lozzi20180,gao2019boosting,colombo2017investigation,cassella2017low,chen2017aluminum,zou2019ultralow,vetury2018high,zuo2011dual,ruby2017deceptively,wang2010fbar,nelson201122muw,wang_film_2020,gao20173,moe2020highly,shealy2017low,chen2019super,shen2019452,kim2021wideband,colombo2025scaln,park202318ghz,park201910}. The other category includes LT, gallium nitride (GaN), and ZnO piezoelectric material systems \cite{takai2019high,popa20132deg,ahmed2021switchable,popa2014band,kubo2003fabrication}. The $Q\cdot k^2$ and $f\cdot Q$ values are calculated from the reported $Q$, $k^2$, and resonance frequency when not explicitly provided in the cited literature.

As shown in Fig.~\ref{fig_12}(a), the proposed bimorph P3F TS-FBAR exhibits high $Q$ across the sub-GHz and low-GHz regimes, maintaining $Q$ above $10^4$ over a broad range of higher-order TS modes and reaching a maximum $Q_s$ of 20,507 at 12~K. Fig.~\ref{fig_12}(b) compares the corresponding FoM. The lower-order TS modes retain competitive FoM owing to the strong coupling enabled by the P3F structure, while the FoM progressively decreases at higher mode orders as $k^2$ decreases. Fig.~\ref{fig_12}(c) compares the $f\!\cdot\!Q$ product, where the proposed device reaches a maximum of $1.98\times10^{13}$~Hz at 1.379~GHz and 12~K. Overall, the comparison highlights the ability of the thick transferred single-crystal P3F LN platform to provide broad frequency scaling while maintaining high $Q$ and competitive $Q\cdotp k^2$ in the sub-GHz and low-GHz regimes.

Future work will focus on identifying and quantifying the dominant loss mechanisms that limit the $Q$ of the bimorph P3F TS-FBAR. The temperature-dependent measurements presented in this work provide an initial basis for separating contributions from intrinsic phonon-related loss, anchor and acoustic-radiation loss, electrode resistive loss, surface and interface loss, and other extrinsic dissipation mechanisms. By combining temperature-dependent characterization with mode-order-dependent measurements and targeted numerical modeling, we can evaluate the relative contribution of each loss channel and identify the dominant limitation to $Q$. Such loss-mechanism studies will guide subsequent optimization of resonator geometry, electrode configuration, material interfaces, and acoustic confinement to further improve $Q$ and $f\!\cdot\!Q$ in overtone resonators.

\section{Conclusion}

This work demonstrates a bimorph P3F LN TS-FBAR that combines a thick transferred single-crystal X-cut LN acoustic cavity with polarization-engineered high-order thickness-shear excitation to support a broad sequence of overtone resonances. The P3F configuration preserves sufficient $k^2$ for a series of TS overtones extending to 1.75~GHz, while the top and floating bottom electrodes enable thickness-field excitation and help confine the acoustic energy within the active region. The fabricated device achieves a peak 3-dB $Q_s$ of 20,507 at 779~MHz and a maximum $f\!\cdot\!Q$ product of approximately $1.98\times10^{13}$~Hz at 1.379~GHz and 12~K. Temperature-dependent measurements further reveal a systematic enhancement in $Q$ at cryogenic temperatures, providing important experimental data for future identification and separation of the fundamental loss mechanisms in P3F LN resonators. These results establish high-order TS excitation in transferred P3F LN as a promising pathway toward compact, frequency-scalable, and low-loss micro-acoustic resonators for sub-GHz and low-GHz frequency-control applications.

\section*{Acknowledgment}
The authors thank Dr. Ming-Huang Li at National Tsing Hua University for helpful discussions.
\balance
\bibliographystyle{IEEEtran}
\bibliography{IEEESettings,references}

\end{document}